\newcommand{\beq}{\begin{equation}}
\newcommand{\eeq}{\end{equation}}
\newcommand{\beqa}{\begin{eqnarray}}
\newcommand{\eeqa}{\end{eqnarray}}
\newcommand{\ket}[1]{| #1 \rangle}

\newcommand{\opa}{\hat{a}}

\newcommand{\opnone}{\hat{n}_1}
\newcommand{\opntwo}{\hat{n}_2}

\documentclass[prl,twocolumn,showpacs]{revtex4}
\begin{document}


\title{Entangled-State Lithography: 
Tailoring any Pattern with a Single State}

\author{Gunnar Bj\"{o}rk}
\thanks{On leave from Department of Electronics, 
Royal Institute of Technology (KTH), Electrum 229, 
SE-164 40 Kista, Sweden}
\email{gunnarb@ele.kth.se}
\homepage{http://www.ele.kth.se/QEO/}
\affiliation{Departamento de \'{O}ptica, 
Facultad de Ciencias F\'{\i}sicas, 
Universidad Complutense, 28040 Madrid, Spain}
\author{Luis L. S\'{a}nchez Soto}
\affiliation{Departamento de \'{O}ptica, 
Facultad de Ciencias F\'{\i}sicas, 
Universidad Complutense, 28040 Madrid, Spain}
\author{Jonas S\"{o}derholm}
\affiliation{Department of Electronics, 
Royal Institute of Technology (KTH), Electrum 229, 
SE-164 40 Kista, Sweden}

\date{\today}

\begin{abstract}
We demonstrate a systematic approach to 
Heisenberg-limited lithographic image formation 
using four-mode reciprocal binominal states. 
By controlling the exposure pattern with a simple 
bank of birefringent plates, any pixel pattern on a 
$(N+1) \times (N+1)$ grid, occupying a square with 
the side half a wavelength long, can be generated 
from a $2 N$-photon state. 
\end{abstract}


\maketitle

When two coherent plane waves of wavelength 
$\lambda$ are made to overlap, a striped intensity
distribution with perfect sinusoidal modulation is formed. 
A simple application of the Rayleigh criterion 
shows then that the minimum resolvable feature 
size occurs at a spacing of  $\lambda/2$, which 
is the best resolution that can be achieved 
classically and is usually known as the diffraction 
limit.

On the contrary, quantum fields allow the 
modulation period to be as small as $\lambda/(2 N)$,  
where $N$ is the average number of photons of the two 
interfering modes.  The origin of the sub-diffraction 
resolution can be envisioned as the photons clustering 
into quasi-particles  with a de Broglie wavelength 
inversely proportional  to the  (relativistic) mass of 
the $N$-photon  quasi-particle~\cite{Jacobson}. 
This is the basis  of the so-called Heisenberg-limited 
interferometry,  that has been experimentally 
demonstrated~\cite{Rarity,Trifonov,Fonseca}. 
In addition, the modulation pattern of two interfering 
quantum  fields is, in general, not sinusoidal, but 
can take many shapes, a fact we shall exploit later.

In this respect, image formation can be seen as a multimode 
generalization of  interferometry. Nowadays, it is recognized 
that for a variety of imaging techniques we can have a 
feature resolution that is ultimately limited not by diffraction, 
but by the quantum fluctuations of the light beams used 
in the experiments~\cite{Fabre}.

In all cases, the existence of the standard diffraction limit 
can be  traced back to the fact that light is treated as a 
classical beam or, which is equivalent, as a stream of  
uncorrelated photons. It has been known for some time 
that entangled photon pairs show unusual resolving 
facts, but only very recently it has been proposed 
the use of  entanglement to increase the resolution indefinitely, 
a fact that opened the possibility of performing quantum optical 
lithography~\cite{Boto}.

In this Letter we go one step further in this promising 
field and discuss the use of reciprocal binomial states 
instead of entangled number states. At first sight, this 
could be seen as a curiosity, but one must take into
account that these states are the basis of the projection 
synthesis method~\cite{Pegg} since they are the 
reference to be used in a beam splitter such that the 
photon counts at the two outputs lead to the experimental 
determination of the expectation value of any projector 
formed by the first $N+1$ number states. Therefore, 
they offer the capability of  tailoring any desired pattern, 
as we shall show below.

A great advantage of our method is that only one entangled
state needs to be generated. All other necessary states 
can be produced from the first by means of a small bank 
of phase plates with a prescribed birefringence. 
Moreover, it also works for any number of photons $N$ 
and offers a very intuitive and simple means of determining 
the  exposure sequence to generate any pixellated pattern 
on a  $(N+1) \times (N+1)$ grid, occupying a square with 
a half  wavelength long side.

Our goal is to establish how we can create arbitrary
2D  patterns on a suitable substrate. Suppose we 
have two counter-propagating beams 1 and 2, in some 
direction we shall denote as the axis $X$, that propagate at 
grazing incidence over a substrate of side $\lambda/2$ 
coated with lithographic resist (in the following we 
will refer to the resist as the film). In general, we have 
to take into account the mode shapes, but when 
restricted to a length $\lambda$ this problem does 
not arise, provided the coherence lengths of the 
wavepackets are much longer than the side of the film. 
In reality, this is a very lax requirement since it means 
that a wavepacket with a coherence length of, say, 
$100  \lambda$  meets this criterion by far. In the 
orthogonal $Y$ direction we have two other beams 
3 and 4,  satisfying the same conditions. The square 
film is  situated in the region where these four beams 
overlap.

If we restrict ourselves to consider $N$-photon states in
the four modes, a quick combinatorial calculation gives  
that there are $(N+1)(N+2)(N+3)/3!$ such pure states. 
In principle, it would be possible to generate this number 
of  distinct patterns (but due to the setup's rotational 
symmetry,  degeneracies would occur). However, 
different nonlinear Hamiltonians would be required to 
generate all the different patterns. In Ref.~\cite{Boto} 
the lithographic film absorption is modeled by an 
$N$-photon absorption process. In this way, higher-order 
interference effects are naturally brought out. Unfortunately, 
the $N$-photon absorption cross section decreases very 
rapidly with increasing $N$ in all materials. However, to 
harness the de Broglie wavelength of $N$-photon 
wavepackets all the photons in the wavepacket must 
collectively interfere and, therefore, $N$-photon absorption 
processes seems be needed.  In addition, for the quantum 
lithography process to make sense, the photosensitive 
``grains'' in the film must be much smaller that the shortest 
de Broglie wavelength encountered in the exposure 
process. Therefore, from the point of view of a 
``grain'', the photon packets in the respective modes 
will be indistinguishable in spite of their different linear 
momenta. Consequently, the  Hamiltonian governing the
absorption can  be written as
\begin{equation}
\hat{H} \propto (\opa_1^\dagger + \opa_2^\dagger
+ \opa_3^\dagger + \opa_4^\dagger)^N
(\opa_1 + \opa_2 + \opa_3 + \opa_4)^N,
\label{eq: Ham}
\end{equation}
where $\opa_i$ is the annihilation operator of mode $i$.

For the moment, let us restrict the problem to one dimension
along the axis $X$. According to Ref.~\cite{Boto}, 
the deposition rate in the substrate $\Delta_N$ is then 
given by the expectation value of the operator 
$e^\dagger{}^N e^N/N!$, where $e = (a_1 + a_2)/\sqrt{2}$. 
But, this is the action of a 50:50 beam splitter for modes 
1 and 2, and inspired by the projection synthesis method, 
it is natural to use  two-mode $N$-photon reciprocal 
binomial states~\cite{Pegg}
\begin{equation}
\ket{\psi^{(N)}} = \frac{1}{\sqrt{\mathcal{N}}}
\sum_{n=0}^N \sqrt{n!(N-n)!} \ket{n,N-n} ,
\label{eq: Relative phase}
\end{equation}
where $\ket{n,N-n} = \ket{n}_1 \otimes 
\ket{N-n}_2$ and $\mathcal{N}=\sum_{n=0}^N n!(N-n)!$ is
a normalization factor. 

Since modes 1 and 2 impinge over the film in 
anti-parallel directions, the accumulated phase of mode 1 
at a distance  $x /\lambda$ from the left edge of the film will be 
\begin{equation} 
\hat{U}=\exp(i k \opnone \lambda x) = 
\exp(i 2 \pi x \opnone) ,
\label{eq: phase}
\end{equation}
where $k=2 \pi/\lambda$ and $\opnone = 
\opa_1^\dagger \opa_1$, while mode 2 will have 
accumulated the phase
\begin{equation} 
\hat{U}=\exp[i k \opntwo \lambda (1-x) ] = 
\exp[i 2 \pi (1-x) \opntwo]  
\label{eq: phase 2}
\end{equation}
at the same location. Using these free-space unitary 
propagation operators, we find that at the location $x$, 
the state (\ref{eq: Relative phase}) is transformed into
\begin{equation}
\ket{\psi^{(N)}_x} = \frac{1}{\sqrt{\mathcal{N}}} 
\sum_{n=0}^N e^{i  2 \pi x(2 n -N)} \sqrt{n!(N-n)!} 
\ket{n,N-n} .
\label{eq: Relative phase x}
\end{equation}
Calculating the pattern deposition rate $\Delta_N$ for
the state $\ket{\psi^{(N)}_x}$, we find that
\begin{equation}
\Delta_N \propto \left 
|\sum_{n=0}^N e^{i  2 \pi x(2 n -N)} \right |^2 
\propto
\frac{\sin^2 [2 (N+1) \pi x]}{(N+1)^2\sin^2(2 \pi x)}.
\label{eq: Absorption}
\end{equation}
We observe that $\Delta_N$ has a highest oscillation period 
in $x$ of $1/[2 (N+1)]$ and an overall periodicity of $1/2$, 
corresponding to the physical lengths $\lambda/[2(N+1)]$ 
and $\lambda/2$, respectively. The former is the shortest 
oscillation period possible using $N$-photon states~\cite{Margolus}. 

Now assume that we translate the substrate a distance 
$\lambda/[4(N+1)]$ to the left, and that we phase-shift mode 1 
by $2 \pi \ell/(N+1)$ $(\ell=1,2,\ldots,N+1)$ relative to mode 2. 
We denote the ensuing state $\ket{\psi_x^{(N,\ell)}}$. 
A simple way of doing this would be to let mode 1 and 2 be 
spatially and temporally degenerate modes, but with 
orthogonal polarizations, see Fig. \ref{fig: setup}. Insertion of 
phase plates with a birefringence corresponding to an optical 
path difference of $\lambda \ell/(N+1)$ and with their principal 
axes parallel to the polarization directions of the modes, would 
provide the needed relative phase shift. The deposition rate 
after these additional modifications are taken into account can 
readily be calculated to be
\begin{equation}
\Delta_N \propto \frac{\sin^2\{[2 (N+1)x-\ell+1/2]\pi \}}
{(N+1)^2\sin^2 \left [ \left (2 x- \frac{\ell-1/2}{N+1} \right ) \pi \right ]}.
\label{eq: Absorption 2}
\end{equation}
Let us now examine the properties of the deposition 
rate function. The function has one peak, roughly 
$1/[2 (N+1)]$ wide in units of $\lambda$, and for 
each successive state $\ket{\psi^{(N,\ell)}}$, where 
$\ell=1,2,\ldots,N+1$, the peak is displaced a distance 
$1/[2 (N+1)]$ to the right. Hence, if we ``pixellate'' the 
substrate along the $X$ direction into $N+1$ pixels, 
each state will expose (or deposit) one pixel, with 
a negligible deposition rate outside the pixel. Note, 
in particular, that the deposition rate due to state 
$\ket{\psi_x^{(N,\ell)}}$ at the center of pixel $\ell_i$ 
is identically zero for all states with $\ell \neq \ell_i$. 
That is, the deposition penalty at the center of each 
pixel is identically zero, independent of which other pixels 
are exposed.

Assume now that we have a similar exposure apparatus 
along the $Y$ direction, i.e., a second synchronized 
generator of the state (\ref{eq: Relative phase}) in modes 
3 and 4, followed by a bank of  birefringent plates, 
a polarizing beam splitter, and rigid mirrors. The state at 
the point $(x,y)$ of the substrate would then be in a 
$2N$-photon four-mode product state 
$\ket{\psi_x^{(N,\ell_x)}} \otimes \ket{\psi_y^{(N, \ell_y)}}$, 
where $\ell_y =1,2, \ldots, N+1$ indicates the differential 
phase shift, in units of $2 \pi/(N+1)$,  between modes 3 and 4. 
The deposition rate is then given by 
\begin{equation}
\Delta_{2N} \propto \left | \sum_{m=0}^N \sum_{n=0}^N 
e^{i  2 \pi (x-\frac{\ell_x-1/2}{N+1})(2 m -N)} 
e^{i  2 \pi (y-\frac{\ell_y-1/2}{N+1})(2 n -N)} 
\right |^2 .
\label{eq: Absorption 3}
\end{equation}
By virtue of being a product state, the deposition rate factors 
into two functions giving our final result
\begin{eqnarray}
\Delta_{2N} & \propto & \frac{1}{(N+1)^4} 
\frac{\sin^2 \{ [2 (N+1)x- \ell_x +1/2] \pi \}}
{\sin^2 \left [ \left (2 x- \frac{\ell_x-1/2}{N+1} 
\right ) \pi \right ]} \nonumber \\
& \times & \frac{\sin^2 \{ [2 (N+1) y - \ell_y + 1/2] \pi \}}
{\sin^2\left [ \left (y-\frac{\ell_y-1/2}{N+1}\right ) \pi \right ]}  .
\label{eq: Absorption 4}
\end{eqnarray}
The price we pay for preparing this $2N$-photon product state, 
rather than a more general state, is that we can only deposit $(N+1)(N+1)$ 
different patterns, instead of close to $(2N+1)(2N+2)(2N+3)/3!$. 
However, the patterns are essentially mutually exclusive, 
so by considering mixed two-mode reciprocal binomial states 
we can built any one of the $2^{(N+1)^2}$ possible different patterns.

In Fig. \ref{fig: serpentine} we have simulated the deposition  of a 
serpentine pattern exposed across the sample. The pixels 
$(2,1)$, $(2,2)$, $(2,3)$, $(2,4)$, $(3,4)$, $(4,4)$, $(5,4)$, $(5,5)$, 
$(5,6)$, and $(5,7)$ have been exposed by preparing a superposition 
of the states with corresponding $\ell_x$ and $\ell_y$. 
Additionally, we have exposed the pixel $(6,4)$ to demonstrate that
nonexposed pixels sitting between two exposed ones, receive
a negligible deposition penalty (identically zero at the pixel center).
In practice, the simplest way  to do this is to expose the 
substrate sequentially and for equal  times, first by one of the states, 
then by the other. The only experimental adjustment that needs 
to be done between the exposures is a rearrangement of the 
bank of birefringent plates. The bank need not be large: to 
accomplish any of the needed $N$ relative phase-differences we 
need only $\log_2 N$ (or actually the smallest integer greater or 
equal to $\log_2 N$) birefringent plates. That is, for a 
$1025 \times 1025$ pixel pattern we need only 10 
birefringent plates in each bank.  The inset shows the 
exposure along  the central ridge. We see that the exposure along 
the ridge has  ripples, but the minimum  is still 89 \% of 
the maximum  exposure. We also see that the exposure penalty 
elsewhere is  small (smaller than 12 \% of the maximum exposure). 
Taking the  nonlinearity, with respect of the exposure dose, of the optical 
density of a lithographic film after development, such a ripple 
is acceptable. We also see one undesirable consequence of
the $\lambda/2$ periodicity of the deposition rate. The pixel
$(5,7)$ will partly expose the pixel $(5,1)$. This problem is fundamental
but can be overcome by sacrificing some of the film area and
considering only the $(N-1)$ central points.

In Fig. \ref{fig: diagonal} we have simulated the deposition  of a 
pattern going diagonally across the sample. This is the worst case 
using the proposed pixellation method. The inset shows the deposition 
rate along the center of  the diagonal ridge. As expected, the 
deposition rate ripple is large, the minima are only 35 \% of the 
maxima. This is unacceptable. To correct this flaw, we shift 
the pattern along both axes by half a pixel. This can be accomplished 
either by translating the substrate by $\lambda/[4(N+1)]$ in both the 
$X$ and  $Y$ directions, or by inserting a birefringent $\lambda/[2(N+1)]$-plate 
at both banks. This allows us to place ``intermediate'' pixels along the 
diagonal,  filling in the deposition rate minima. In Fig. \ref{fig: filled in diagonal} 
we have simulated the same pattern by exposing pixels $(2,1)$, $(2,2)$, and 
$(5,7)$ with a full exposure dose, pixels $(2,3)$ and $(5,6)$ with a 
relative exposure dose of 0.83, pixels $(3,4)$ and $(4,5)$ with a 
relative exposure dose of 0.66, and added intermediate pixels at pixel 
positions $(2.5,3.5)$, $(3.5,4.5)$, and $(4.5,5.5)$, at a relative exposure 
dose of 0.66. The figure inset shows the deposition rate along the 
diagonal ridge, demonstrating that very good patterns can 
be formed in this way. The ridge's deposition rate maximum is 1.04, 
and the minimum is 0.90 of the deposition rate of an isolated 
pixel exposed with the nominal dose.

It is worth discussing what the experimental challenges 
with quantum lithography are. We see three main hurdles. 
The first is to develop a reasonably sensitive $N$-photon absorption 
film. Such a film seems to be the only way to fully capitalize on the 
de Broglie wavelength of photon packets, and our proposal 
will share this difficulty with any other proposal. 
The second hurdle is how to generate the reciprocal binomial states. 
Today, no generator of such a state with $N >2$ exists, while for
$N=2$ a two-photon source and an appropriate beam splitter can
be used. However, at least one proposal to do this job,
using  atom-photon interaction, is at hand~\cite{Moussa}. 
In, e.g., the proposal of Boto \textit{et al.}~\cite{Boto}, 
each pattern needs a different, and typically 
equally complicated, state. Unfortunately, to make a generator of 
an infinite (or at least large) set of different entangled states seems 
impossible  with present technology. Thirdly, the exposed area will be very 
small in comparison to the wafers, and even the chips, presently 
manufactured in the semiconductor industry. However, using 
only two modes along each direction will always produce periodic 
patterns. To increase the spatial repetition period, while keeping 
the resolution, more modes are needed. This is not impossible, but 
at least in our opinion, one must first establish a procedure for how 
to generate arbitrary patterns in a systematic manner. We hope 
this paper can serve as a basis for this line of development.

\begin{acknowledgments}
This work was supported by the Swedish Research Council 
for Engineering Sciences (TFR), the Swedish Foundation for 
Strategic Research (SSF) and the Swedish Natural Science 
Research Council (NFR).
\end{acknowledgments}


\begin{thebibliography}{99}

\bibitem{Jacobson} 
J. Jacobson, G. Bj\"{o}rk, I. Chuang, and Y. Yamamoto, 
\prl \textbf{74}, 4835 (1995).

\bibitem{Rarity} 
J. G. Rarity and P. R. Tapster, 
\pra \textbf{41}, 5139 (1990).

\bibitem{Trifonov} 
A. Trifonov, T. Tsegaye, G. Bj\"ork, 
J. S\"oderholm, E. Goobar, M. Atat\"ure, and A. V. Sergienko, 
J. Opt. B: Quantum Semiclass. Opt.  \textbf{2}, 105 (2000).

\bibitem{Fonseca} 
E. J. S. Fonseca, C. H. Monken, and S. P\'{a}dua, 
\prl \textbf{82}, 2868 (1999).

\bibitem{Fabre}
M. I. Kolobov and C. Fabre,
\prl \textbf{85}, 3789 (2000).


\bibitem{Boto} 
A. N. Boto, P. Kok, D. S. Abrams, 
S. L. Braunstein, C. P. Williams, and J. P. Dowling, 
\prl \textbf{85}, 2733 (2000).

\bibitem{Pegg} 
 S. M. Barnett and D. T. Pegg,
\prl \textbf{76}, 4148 (1996); 
D. T. Pegg, S. M. Barnett, and L. S. Phillips, 
J. Mod. Opt. {\bf 44}, 2135 (1997).



\bibitem{Margolus} 
N. Margolus and L. B. Levitin, 
Physica D \textbf{120}, 188 (1998); 
J. S\" oderholm, G. Bj\" ork, T. Tsegaye, and A.Trifonov, 
\pra {\bf 59}, 1788 (1999).

\bibitem{Moussa}  
M. H. Y. Moussa and B. S. O. Baseia, 
Phys. Lett. {\bf A238}, 223 (1998).

\end{thebibliography}

\newpage

\begin{figure}
\caption{A schematic showing the proposed setup to expose the film in the 
$X$ direction. PBS denotes a polarization beam splitter. It is assumed that 
a synchronous state generator, a second birefringent phase-plate bank, 
and appropriate mirrors are also arranged in the $Y$ direction.}
\label{fig: setup}
\end{figure}

\begin{figure}
\caption{Calculated deposition rate for a serpentine pattern 
made from a two six-photon two-mode reciprocal binomial states. 
We have also deposited in the pixel $(6,4)$. The inset shows 
the deposition profile along the line crossing the center of pixels 
$(5,1)$ and $(5,7)$.}
\label{fig: serpentine}
\end{figure}

\begin{figure}
\caption{Calculated deposition rate for a S-curve made from  two 
six-photon two-mode relative phase states. The inset shows the 
deposition profile along the diagonal, hilly ridge. 
The profile goes from pixel $(1,2)$ to pixel $(6,7)$.}
\label{fig: diagonal}
\end{figure}

\begin{figure}
\caption{An improved deposition rate function, where intermediate pixels 
have been placed to make the diagonal ridge smoother. The inset shows 
the deposition along the ridge. The result is a substantial improvement from 
the previous figure.}
\label{fig: filled in diagonal}
\end{figure}

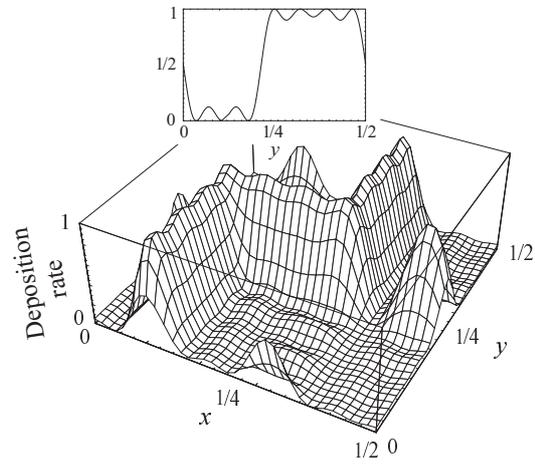
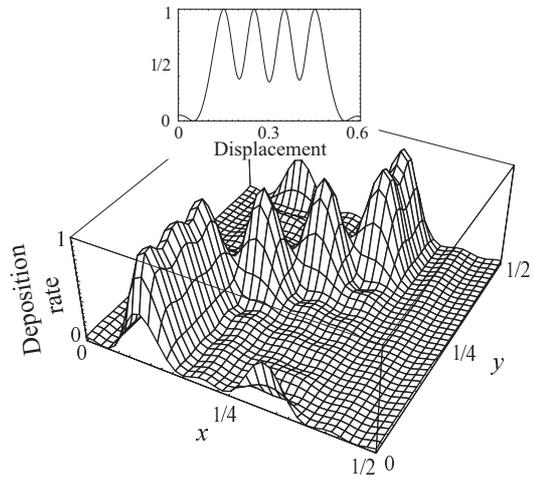
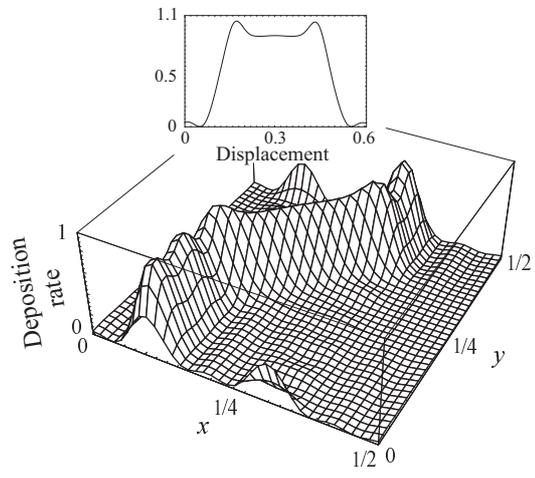
\end{document}